# Scaler rates from the Pierre Auger Observatory: a new proxy of solar activity

The Pierre Auger Collaboration,[1] I. Bizzarri,[2] C. Dionese,[2] and S. Mancuso[3]

[1] Observatorio Pierre Auger, Av. San Martín Norte 304, 5613 Malargüe, Argentina
[2] Dipartimento di Fisica, Università degli Studi di Torino, Via Pietro Giuria 1, Torino, Italy
[3] INAF, Osservatorio Astrofisico di Torino, via Osservatorio 20, Pino Torinese 10025, Italy

## ABSTRACT

The modulation of low-energy galactic cosmic rays reflects interplanetary magnetic field variations and can provide useful information on solar activity. An array of ground-surface detectors can reveal the secondary particles, which originate from the interaction of cosmic rays with the atmosphere. In this work, we present an investigation of the low-threshold rate (scaler) time series recorded in 16 years of operation by the Pierre Auger Observatory surface detectors in Malargüe, Argentina. Through an advanced spectral analysis, we detected highly statistically significant variations in the time series with periods ranging from the decadal to the daily scale. We investigate their origin, revealing a direct connection with solar variability. Thanks to their intrinsic very low noise level, the Auger scalers allow a thorough and detailed investigation of the galactic cosmic-ray flux variations in the heliosphere at different timescales and can, therefore, be considered a new proxy of solar variability.

## 1. INTRODUCTION

During their propagation through the heliosphere, Galactic Cosmic Rays (GCRs) interact with the solar wind and the heliospheric magnetic field, which modify their energy spectra. Changes in the interplanetary medium related to variations in the Sun's activity and to solar transient events thus determine the magnetic deflection of the trajectories of GCR particles, modifying the flux of the GCRs reaching the Earth's atmosphere.

The process through which GCR particles interact with magnetic irregularities in the solar wind can be described as diffusion combined with convection and adiabatic energy losses (Parker 1965). In particular, during the minimum phase of solar activity, when the Sun is quiet, GCRs have a maximum intensity at Earth, and vice versa during solar maximum conditions, so solar activity effectively modulates periodically the GCR flux with the same solar decadal cycle. Apart from this long-term modulation associated with the solar cycle, short-term variations of the flux of GCRs are also produced by the perturbed interplanetary condition near the Earth, such as interplanetary coronal mass ejections (ICMEs) (e.g., Richardson & Cane 2010) or stream interaction regions (SIRs) (e.g., Richardson 2018). These temporal depressions in the GCR flux, generally known as Forbush decreases (Forbush 1937), are a consequence of changes on the GCRs transport plasma properties. While ICMEs are manifestations of solar eruptions, SIRs arise when fast solar wind flow (originating in the core of coronal holes) reaches slow solar wind flow (Grieder 2001). Since the solar wind velocity is radial from the Sun, this interaction is only possible because of the solar rotation. When a coronal hole remains during more than one solar rotation, the associated recurrent SIR is called a corotation interaction region (CIR) (Lockwood 1971; Cane 2000; Richardson 2004; Dumbović et al. 2011; Richardson 2018; Vršnak et al. 2022).

The Pierre Auger Observatory (Pierre Auger Collaboration 2015) is the largest cosmic-ray observatory to date, specifically designed to study the physics of cosmic rays at the highest energies, above $3\times10^{17}$ eV. It includes observa-

Corresponding author:
spokespersons@auger.org

tions of the fluorescence light produced by air-shower secondary particles as they propagate through the atmosphere and the direct measurement of secondary particles reaching ground level. Since 2005, the Observatory has also recorded the low-threshold rates (scalers) corresponding to signals with energy between 15 Mev and 100 MeV revealed by all the water-Cherenkov surface detectors of the array with a methodology known as the single-particle technique (Morello et al. 1984). Apart from allowing monitoring the long-term stability of the detectors, scalers can be used for searching for transient events, such as gamma-ray bursts (Bertou 2008; Pierre Auger Collaboration 2009), solar flares (Abbasi et al. 2008), and Forbush decreases (Pierre Auger Collaboration 2011; Dasso et al. 2012), which are expected to produce coherent variations in their counting rates, and for investigating long-term trends in the heliospheric modulation of GCRs during the solar cycle (Schimassek 2020).

In this work, we show that Auger scaler data are not only complementary to those provided by neutron monitors or muon detectors but, thanks to the very low noise level (resulting from statistical fluctuations intrinsic to the original signal or added by the measurement), and to the higher statistical significance related to the very high count rates ($\sim 10^6$ counts per second), they allow for a thorough and detailed investigation of the GCR flux variations in the heliosphere. Besides the imprint of the decadal solar cycle previously shown (e.g., Schimassek 2020), we reveal here GCR variations from the annual to the daily scale. Through the spectral analysis of a uniformly resampled 16-year-long scaler time series obtained by applying an Auto-Regressive (AR) gap-filling technique to fill several gaps in the time series, we extract from the noise the significant oscillatory components of the time series and reconstruct their time evolution. Moreover, we conduct an in-depth investigation to understand the phenomena at the origin of the detected oscillations, particularly the possible relationship with the solar modulation over different time scales.

After introducing the scaler rate time series in Section 2, the results obtained regarding the spectral content are shown in Section 3. The results are finally discussed in the concluding Section 4.

## 2. SCALER RATE AT THE PIERRE AUGER OBSERVATORY

### 2.1. *The Pierre Auger Observatory*

The Pierre Auger Observatory (Pierre Auger Collaboration 2015) is located at an altitude of 1400 m above sea level in the Argentinian Pampa Amarilla. It is a hybrid system, completed in 2008, that combines surface detectors (SDs), which also measure the scaler rate, and fluorescence telescopes (FDs).

The SDs form an array consisting of 1600 water-Cherenkov detectors (Allekotte et al. 2008) arranged on a 1500 m triangular grid that covers about 3000 $\mathrm{km}^2$. Further 60 water-Cherenkov detectors, with a 750 m spacing, form a 27 $\mathrm{km}^2$ infill region, allowing for extension to lower energies. In each surface detector, Cherenkov radiation produced by the shower particles passing through the water volume is measured using three photomultiplier tubes (PMTs). The amount of Cherenkov radiation is measured in units of charge produced by a vertical through-going muon (Bertou et al. 2006).

### 2.2. *The scaler rates time series*

In 2005, a single-particle technique mode (Morello et al. 1984) was implemented for the full array of SD detectors consisting of recording the rate of signals above a low threshold, the "scaler mode". The Auger scalers record the counting rates of signals within the range [4, 20] analogue-to-digital converter (ADC) counts above baseline, approximately corresponding to the deposited energy range [15, 100] MeV, resulting in an average total count rate of $\sim 3\times 10^6$ per second or $\sim 2000\,\mathrm{s}^{-1}$ per Cherenkov detector. The dominant contribution to the count rate comes from cosmic rays of energies between 10 GeV and a few TeV primary energy (Dasso et al. 2012).

Scaler data for each detector are stored every second. Measured scaler rates are affected by several factors, such as atmospheric conditions, the intrinsically non-constant rate of low-energy particles, and, eventually, instrumental instabilities. Therefore, before looking for transient events and studying long-term solar modulations, the Auger scaler rates must be treated and corrected, as in previous analyses of these data (Masías-Meza 2015; Schimassek 2020), thus obtaining the so-called corrected scaler rate $\Gamma_i^{(c)}$ of station $i$.

The most important condition to ensure high data quality is the stable operation of all three PMTs of a water Cherenkov detector because of the three-fold coincidence condition used for the scaler trigger. We used the PMT-selection to reconstruct ultra-high-energy cosmic rays (Pierre Auger Collaboration 2020) and select high-quality data. Nonetheless, there are remaining instabilities in the scaler data, for example, due to the effect of thunderstorms (Schimassek 2020), that must be removed to obtain a high-quality data sample. Therefore, we employ a median and median-absolute-deviation-based removal of water Cherenkov detectors if their measured rate is significantly above its median and no more than two are affected in each second.

The data series must also be corrected for atmospheric pressure and detector ageing before being used in long-term analyses. The pressure correction is derived from the correlation analysis of the measured rate with the observed atmospheric pressure at the site as in previous work (Pierre Auger Collaboration 2011; Schimassek 2020) and applied as a multiplicative correction. No correction with temperature is applied as no significant correlation of the scaler rate with ground temperature is found (Pierre Auger Collaboration 2011).

The detector ageing influences the signal shape and the measured scaler rate. The signal shape can be monitored using the ratio of the average charge (area) of a vertical-muon with its associated average maximal signal (peak) (Abreu et al. 2011). We use the "area-over-peak ratio" to correct the scaler rate for the ageing by applying a multiplicative correction as in Masías-Meza (2015) and Schimassek (2020).

For each station, the rate measured within a specific five-minute time interval is given by the arithmetic mean over all the seconds in the considered interval not removed by one of the selection criteria.

Finally, the scaler rate value associated with such a time interval is obtained by averaging all stations. Since the stations are deployed at different altitudes, ages, and PMT gains, the scaler rate is not identical between stations. To compensate for these differences, the corrected scaler rate $\Gamma_i^{(c)}$ of a station $i$ is scaled by the reference value $\langle\Gamma_i\rangle$, obtaining the relative scaler rate

$$r_i(t) = \frac{\Gamma_i^{(c)}(t)}{\langle\Gamma_i\rangle}. \tag{1}$$

The reference value $\langle\Gamma_i\rangle$ is the mean count rate of the station $i$ during the year 2013, which is roughly in the middle of the dataset. The idea behind this scaling is that the relative response of the stations to changes in the physical rate is more consistent than the absolute count rate. Furthermore, it is a way to make the response of the whole SD array independent of the number of currently working stations.

Finally, we apply a correction for the non-integer part of the drifting baseline that determines the integer threshold of the scaler trigger. This correction reduces the observed fluctuations on time scales smaller than one day by parameterizing the dependence of the observed counts per fractional baseline drift. This correction leads to a minor offset from unity for the rate used for normalization in 2013 due to the correction to the integer value (3 ADC above baseline) instead of the average (3.5 ADC above baseline) that was not taken into account in the normalization.

Figure 1 shows the time series resulting from the mean of the relative scaler rate $r_i(t)$ over all the stations.

Specifically, Figure 1(a) shows the relative scaler rate sampled at 12 h intervals, characterized by occasional gaps resulting from applied selections, including a few short gaps (of the order of days), along with longer ones that occurred in 2007 (37 d), in 2012 (24 d), in 2013 (29 d) and 2014 (19 d). These widest gaps are highlighted in green.

To obtain a uniform series necessary for the analysis techniques described below (see Section 3), all gaps were filled using a gap-filling method based on autoregressive (AR) models, with the order determined through the Akaike criterion (Akaike 1969).

Using this method, missing data points are substituted with estimates derived from the remaining data samples' forward and backward AR fits. The maximum number of samples used in the estimation and the order of the AR model were chosen to lead to the most accurate reconstruction of randomly selected periods of similar length in the existing data series. Finally, the gap-filled points were computed by averaging 100 gap-filling estimates, each based on optimized parameters for the AR model, since the set of artificial gaps is randomly created in each iteration. The resulting (filled) series covers more than 16 years, from 01 January 2006 to 19 March 2022, for 11 844 data points. The end of the time series in March 2022 is determined by the start of significant deployment of the upgrade of the Pierre-Auger observatory, AugerPrime (Pierre Auger Collaboration 2016), which includes a change of the station electronics (Halim et al. 2023). This series shows a decadal modulation as well as an annual oscillation.

## 3. SPECTRAL ANALYSIS AND RESULTS

Appropriate methods for spectral analysis are necessary when dealing with the potential presence of non-sinusoidal variations in time series and with spectra displaying widely varying power ranges encompassing both weak and strong spectral components. Consequently, an accurate spectral analysis has been performed using advanced spectral methods to reliably reveal the significant periodic components within the relative scaler rate series $r_i(t)$. Singular Spectrum Analysis (SSA; see Appendix A) is a spectral method to confidently extract deterministic components from the series. In contrast to the classical Fourier spectral methods, which use a fixed basis of harmonic functions (sines and cosines), SSA takes advantage of data-adaptive basis functions; this feature makes the method particularly useful also for short and noisy time series. This method decomposes a time series into statistically independent components that can be categorized as oscillatory patterns or noise. The extracted periodic components can exhibit modulations in terms of amplitude and phase. The statistically significant components were detected and extracted from the background noise using a recursive SSA method based on a Monte Carlo test (MC-SSA). For further details, see Appendix A. The noise in the series results from statistical fluctuations intrinsic to the original signal or added by the measurement.

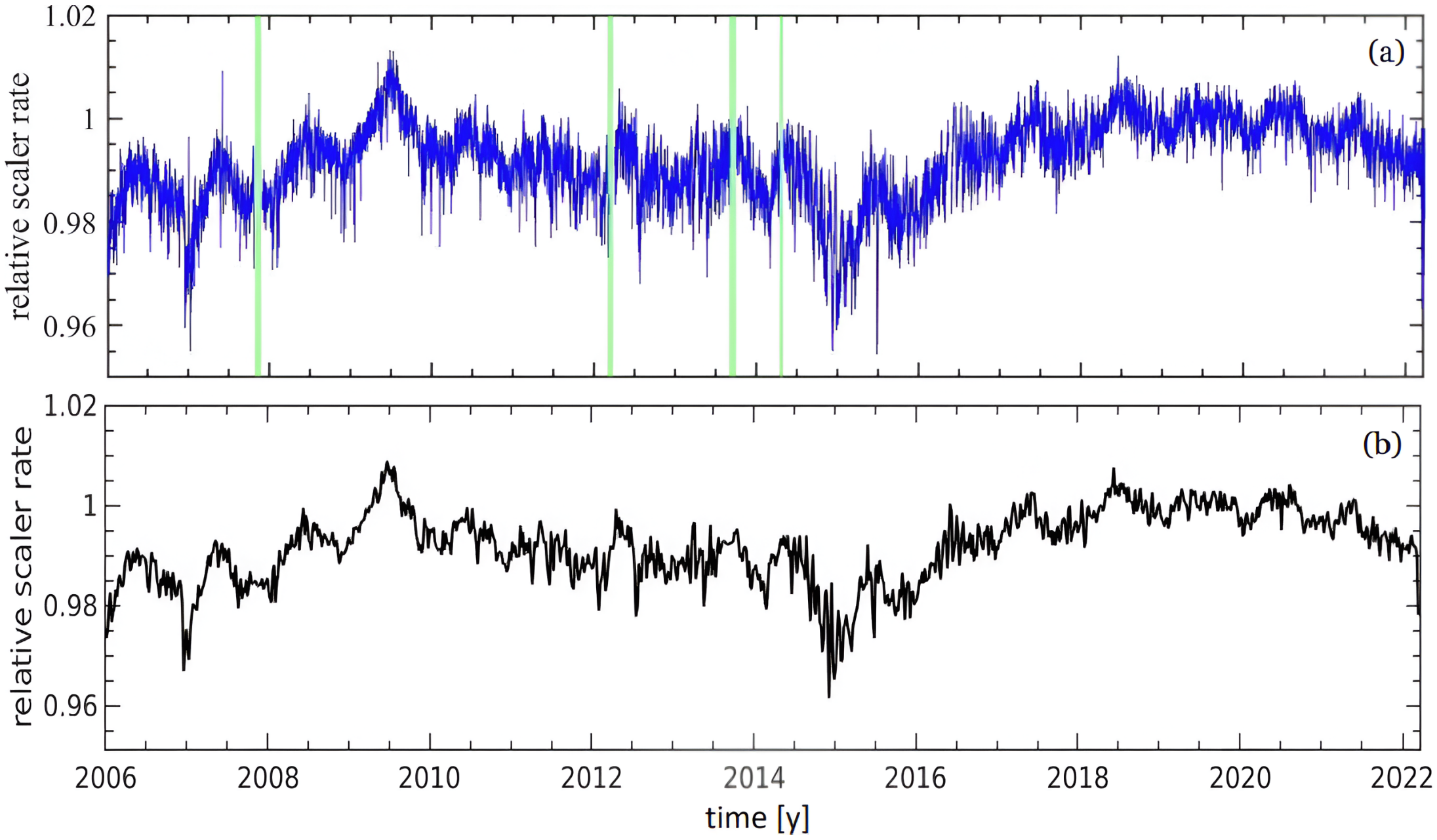

**Figure 1.** Relative scaler rates series from 01 January 2006 to 19 March 2022. The rate incorporates all the corrections detailed in Schimassek (2020) and the text. Panel (a) displays the series sampled at 12-hour intervals. Some gaps in the series have been filled through a gap-filling process relying on an Auto-Regressive model (see text). The most substantial gaps are marked in green. The scaler rate in panel (b) was obtained by resampling the original series every 6 days after applying the gap-filling procedure to the series in panel (a).

The analysis was applied to a downsampled version to extract the periodic oscillations present in the time series, uncovering the oscillations corresponding to periods greater than a few days. This downsampling was done to reduce computational requirements while preserving adequate time resolution. In detail, starting from the gap-filled version of the 12-hour-sampled series, a 6-day-sampled series has been obtained (987 points), thus reducing the length of the series under analysis to one-twelfth of its original size.

The series is shown in panel (b) of Figure 1. The adopted window length $M$ was set equal to 150 samples, corresponding to a window width $W = M\,\Delta t \approx 2.5\,\mathrm{y}$, where $\Delta t = 6\,\mathrm{d}$ is the sampling interval. The robustness of the results was tested using a wide range of $M$ values. The spectrum of the series is represented in terms of power vs frequency in Figure 2.

We use a recursive Monte Carlo (MC) method (Allen & Smith 1996) that reliably identifies the spectral components in a time series. Starting from a null hypothesis of pure red noise, represented by a first-order autoregressive process AR(1), the model is made iteratively more complex until the data cannot statistically reject the model. The method creates 10,000 MC data samples per iteration to obtain a band in frequency space at a given confidence level. The hypothesis test compares the SSA spectrum of the original series with the Monte Carlo band.

The first AR(1) noise assumption is usually used since a large class of physical processes generates series with larger power at lower frequencies. This is done to avoid overestimating the predictability of the system by underestimating the amplitude of the stochastic component of the time series (Vautard et al. 1992).

The model explaining the series includes the following significant components at 99% c.l.: a decadal trend, an annual oscillation, and modes of variability with periods of ~9 month, ~6 month, ~28 d, ~20 d, and ~14 d. The gray bars in Figure 2 constitute the Monte Carlo band. As one can see, no anomalous power exceeds this band except those corresponding to the significant components mentioned above and highlighted by the red squares. The black dots indicate the spectral components that can be parameterized as red noise or are not significant.

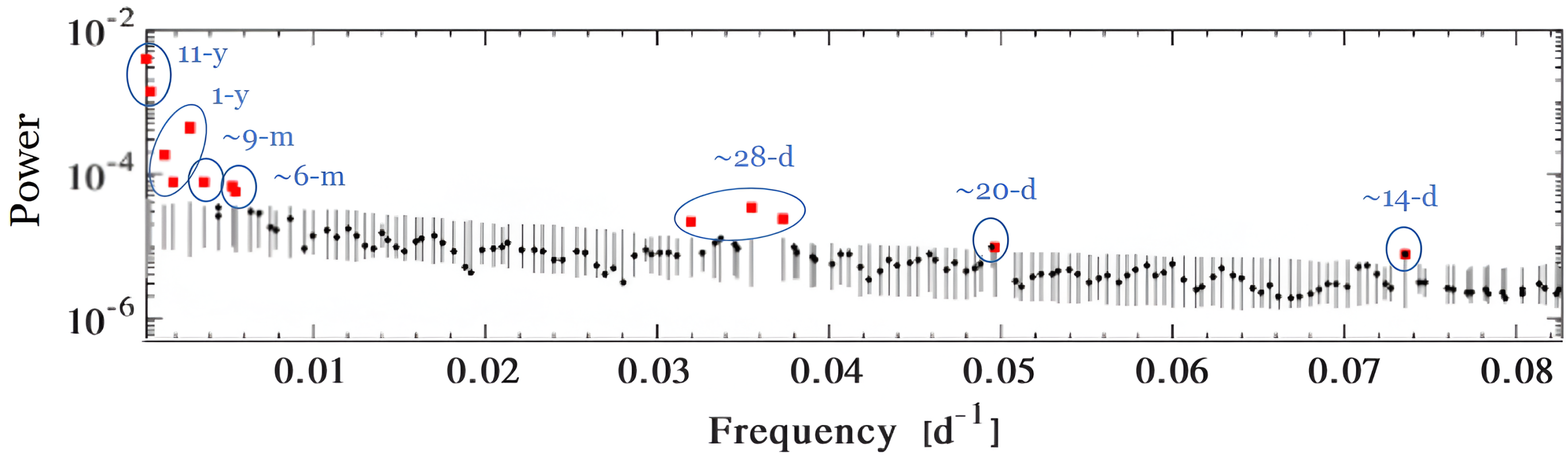

**Figure 2.** MC-SSA spectrum of the relative scaler rate. The Monte Carlo ensemble size is 10 000. The gray bars, which bracket 99% of the power values obtained from the ensemble, represent the Monte Carlo band. The significant spectral components are indicated by the red squares, while the black dots represent the spectral components that can be parameterized as red noise. The significant components with the same period specified in blue are grouped with blue boundaries.

Figure 3 shows the significant components revealed by MC-SSA analysis in the relative scaler rate series (a), with decadal scale (b), annual (c), ~9 month (d), ~6 month (e), ~28 d (f), ~20 d (g), and ~14 d (h) periods. The percentage of variance described by each component is also indicated within each panel. The total variance corresponding to all the significant components is about 88%. Therefore, we point out that the noise level in the scalers series is extremely low (about 12%). For comparison, the bottom panel of Figure 3 shows the sunspot number (SN) series with a 6 d resolution. The SN is an index that quantifies the abundance of lower-temperature "spots" associated with regions of high magnetic field strength on the Sun's surface and is the best-known proxy of solar activity: higher SNs indicate higher solar activity. The used SN time series has the same sampling interval of 6 d as the scaler rates and was derived by resampling the daily data currently maintained by the World Data Center Sunspot Index and Long-term Solar Observations (SILSO World Data Center 2021). Similar periodicities have been found in neutron monitor data and various solar indices (Singh & Badruddin 2019; López-Comazzi & Blanco 2022, 2020) (see Sections 3.1,3.2,3.3,and 3.4).

### 3.1. *Decadal variability*

The Sun is a magnetically active star whose activity depends on the magnetic dynamo process operating in the solar convection zone at a depth of about 200 Mm. Solar magnetic activity is visible through the cyclical appearance of sunspots and active regions on the Sun's surface, varying on different timescales. The main cycle (called the Schwabe cycle) has a period of about 11 y and leads to polarity inversion of the global solar magnetic field.

The decadal modulation, shown in Figure 3(b), is linked to the Schwabe solar cycle and describes most of the signal variance (~68%). In Figure 4(a), this component is directly compared to the decadal one (red curve) revealed by SSA in the SN series (light red curve) using the same window length $M$ adopted for the scalers analysis. The anti-correlation between the two decadal components is revealed (correlation coefficient $r = -0.62$, $p_{\text{value}} < 10^{-5}$). However, a phase displacement between the scalers and the SN decadal components, which varies along the series, is visible. For instance, a lag of about eight months is observed around 2009, which decreases in time, reaching a value of about three months around 2013. The delay in the period 2014 to 2015 is about one year. Instead, the maximum around 2019 is almost in phase with the minimum observed in the SN cycle. A time lag between GCR intensity and solar activity, particularly concerning long-term modulations, has been discussed in several papers (López-Comazzi & Blanco 2022; Ross & Chaplin 2019; Singh et al. 2008). It is generally related to the above-mentioned reversal of the solar magnetic polarity. The polarity cycle is defined as positive ($A > 0$) when the Northern magnetic field is directed away from the Sun and as negative ($A < 0$) when it is pointed toward the Sun. Differences in the behavior of charged particles among polarity cycles occur because, during the $A > 0$ cycle, positively charged particles tend to drift toward the Sun along the polar regions. In contrast, electrons mainly drift along the equatorial heliospheric current sheet, which divides the heliosphere's two oppositely oriented magnetic-polarity hemispheres. When the polarity cycle switches ($A < 0$), the opposite behavior occurs, and when protons drift inward mainly through the equatorial regions of the heliosphere, they encounter the equatorial heliospheric current sheet and are progressively reduced by its increasing waviness as solar activity grows: the wavy heliospheric current sheet thus has significant physical effects in cosmic ray modulation. This produces maxima in the cosmic-ray intensity profiles, which are wider when $A > 0$ (see, e.g., Potgieter 2013 for an extended review). This behavior is reflected in the decadal scaler cycle (black line in Figure 4), which shows a maximum around 2019 wider than the one around 2009. This could cause the observed phase displacement, which varies between the two periods of opposite magnetic polarity. Therefore, a 22 y cycle could be present in the scalers (corresponding to the Hale cycle of solar activity), which is, however, not detectable in the spectral analysis due to the limited time interval covered by the data. The peak around 2009 has already been reported in Masías-Meza (2015) using counting rates in two different energy ranges.

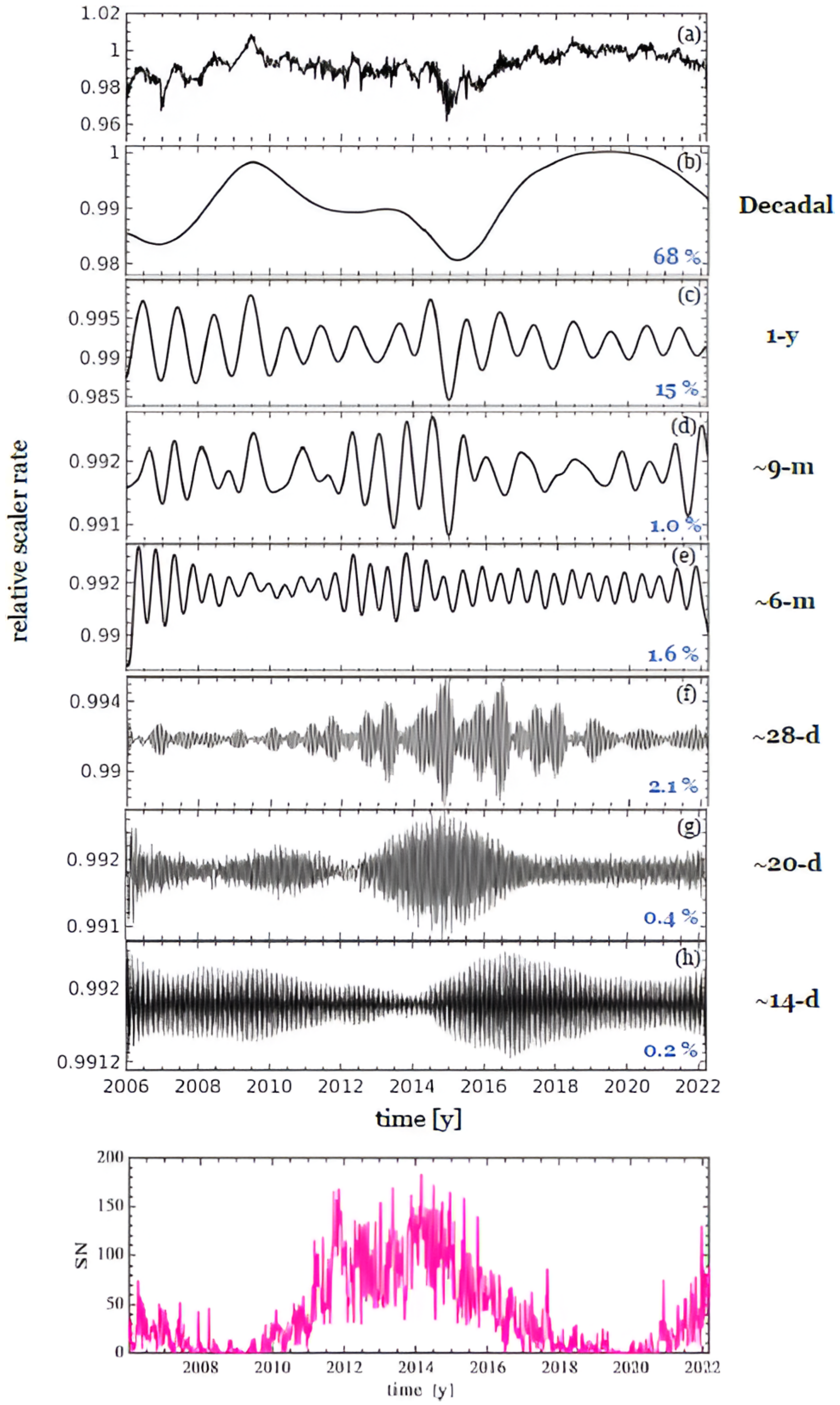

**Figure 3.** The significant components identified through MC-SSA analysis in the relative scaler rate series (a), with decadal scale (b), annual (c), ~9 month (d), ~6 month (e), ~28 d (f), ~20 d (g), and ~14 d (h) periods. The variance described by each component is also indicated in per cent inside each panel. The sunspot number (SN) series sampled every 6 days is also shown in the bottom panel.

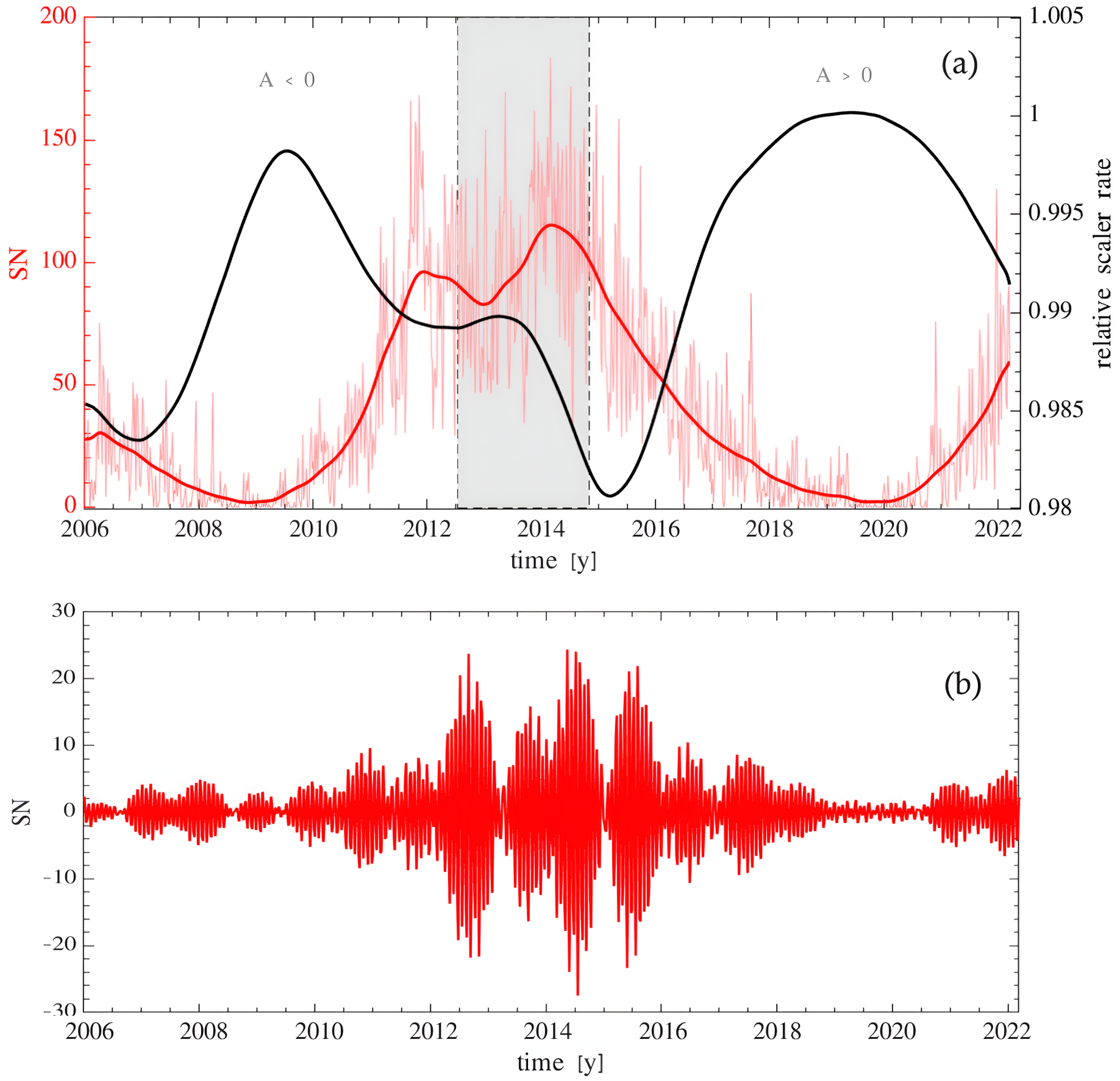

**Figure 4.** (a) Comparison between the decadal trend revealed in the Auger scaler rate (black curve) and the SN series sampled every 6 d (shaded red curve), superimposed by the decadal modulation revealed in the latter by SSA (red curve). An anticorrelation among the decadal trends is visible. The shaded gray bar represents the total time interval required for the polar field reversal in both hemispheres from June 2012 to November 2014. Panel (b) shows the 28 d oscillation revealed in the SN time series.

### 3.2. *Annual variability*

The annual oscillation (see Figure 3(c)) shows minima at the beginning of each year (December to January) and maxima in the middle (June to July). The origin of this annual modulation can be related to a combination of different factors.

Among the terrestrial causes is the seasonal variation of the atmospheric temperature due to the inclination of the Earth's axis relative to the ecliptic plane. The upper atmosphere is cyclically affected by seasonal temperature variations that alter the mean free path of muons at the relevant production heights (Castagnoli & Dodero 1967; Andreyev et al. 1990; Ambrosio et al. 2003; Aglietta et al. 2008). Specifically, in December-January, when the temperature in Malargüe is higher, the expansion of the atmosphere increases the path from the generation point to the surface detectors, leading to a higher probability of muon decay and consequently to a lower muon flux at the ground.

More work on the fraction of the muonic signal in the scalers and on understanding other systematic effects due to seasonal changes in the atmosphere is necessary before firm conclusions can be reached.

Among the extraterrestrial causes of the annual scaler-rate cycle, there is the variation of the Earth-Sun distance due to the eccentricity of the Earth's orbit and to the effect of the asymmetry of the heliospheric magnetic field (see, e.g., Barker & Hatton 1971; Nagashima et al. 1998; Jeong & Oh 2022).

### 3.3. *6-month and 9-month variability*

Two strong components are identified through the MC-SSA analysis in the relative scaler rate series, namely, ~9 month and ~6 month oscillations (see Figure 3(d), (e)). To investigate the solar origin of these two components, we analyzed the total sunspot area (SA) records, acquired daily by the USAF/SOON telescopes[1] with the contribution of the US NOAA. The respective contributions of the two solar hemispheres to the total sunspot area were determined by separately analyzing the SA time series of the Northern (NH) and Southern (SH) hemispheres.

In Figure 5, we show the full Sun (total) SA series (panel b) together with the NH (panel a) and SH (panel c) hemispheric series (black lines). The most prominent peaks in the full Sun series, occurring at the end of 2011 and during the period 2014 to 2015, are related to the different activities of the two hemispheres, as highlighted by the shaded red and blue bars. We applied the MC-SSA to the three resampled (with 6 d time resolution) series, using the same window length $M = 150$ adopted for the Auger scalers analysis.

In Table 3.3, we show the percentage of the total variance associated with the SSA significant components of the three SA series and the scalers series. The last two rows show the total variance related to signal and noise for each series. We notice that the total SA series shows the same spectral content as the scalers series, except for the 14 d and the annual components. The latter is missing, as expected, due to its seasonal origin. Furthermore, analysis of the two hemispheric SA series reveals that the ~6 month and ~9 month periodicities are due to the NH and SH, respectively.

The two reconstructed monthly components are shown in Figure 5(a) (red curve) and Figure 5(b) (blue curve), respectively, showing a higher variability in correspondence to the two main peaks. It is important to point out that the noise level of the scalers series is lower by a factor greater than 2 compared to the total SA series and by a factor of about 3 compared to the NH and SH series.

The ~6 and ~9 month variability is related to the known solar Rieger-type periodicity (Rieger et al. 1984), which was initially attributed to the 154 d periodicity in gamma-ray flares observed by the Solar Maximum Mission (SMM) near the maximum of Solar Cycle 21. The analysis of different indicators of solar magnetic activity during the past few cycles, namely, X-ray flares (Dennis 1985; Bai & Sturrock 1987; Kile & Cliver 1991; Dimitropoulou et al. 2008), sunspot group numbers (Lean 1990; Carbonell & Ballester 1990, 1992; Oliver et al. 1998; Ballester et al. 1999), 10.7 cm radio flux and SN (Lean & Brueckner 1989), occurrence rates of solar flare energetic electrons (Droege et al. 1990), type II, III, and IV radio bursts (Verma et al. 1991; Lobzin et al. 2012), as well as microwave (Kile & Cliver 1991) and proton (Bai & Cliver 1990) flares, confirmed the existence of such a periodicity. This component was found to be strong in some cycles and weak or lacking in others. Moreover, its period is cycle-dependent: the stronger the solar cycle, the shorter the period (Gurgenashvili et al. 2016).

The physical reason for the occurrence of Rieger-type periodicities has been debated for decades, and different mechanisms have been suggested to explain the enigmatic features of this component (Ichimoto et al. 1985; Bai & Sturrock 1991; Lou 2000; Sturrock et al. 2013; Sturrock et al. 2015). On the other hand, Rieger-type periodicities usually appear near solar-cycle maxima.

Recent studies show that this periodicity is probably related to perturbations in the solar internal dynamo layer, where the large-scale magnetic field is generated, attributable to locally generated magneto-Rossby waves (Zaqarashvili et al. 2010). In fact, variations in the differential rotation and magnetic field strength throughout the solar cycle can enhance the growth rate of particular harmonics of magnetic Rossby waves in the upper part of the tachocline (the transition layer between the radiative interior and the outer convective zone), especially around the maximum of the solar cycle.

In turn, the generation of these unstable harmonics may lead to the periodic emergence of magnetic flux at the solar surface due to magnetic buoyancy, thus modulating the ICMEs events and consequently the GCR flux at these time scales. Finally, since the dispersion relation of magnetic Rossby waves depends on the unperturbed magnetic field strength (Zaqarashvili et al. 2007), the amplitude of the specific enhanced harmonics will differ in the North and South hemisphere depending on their level of activity (Gachechiladze et al. 2019), thus leading to the observed North-South asymmetry in Rieger periodicities shown in Figure 5.

[1] http://solarcyclescience.com/activeregions.html

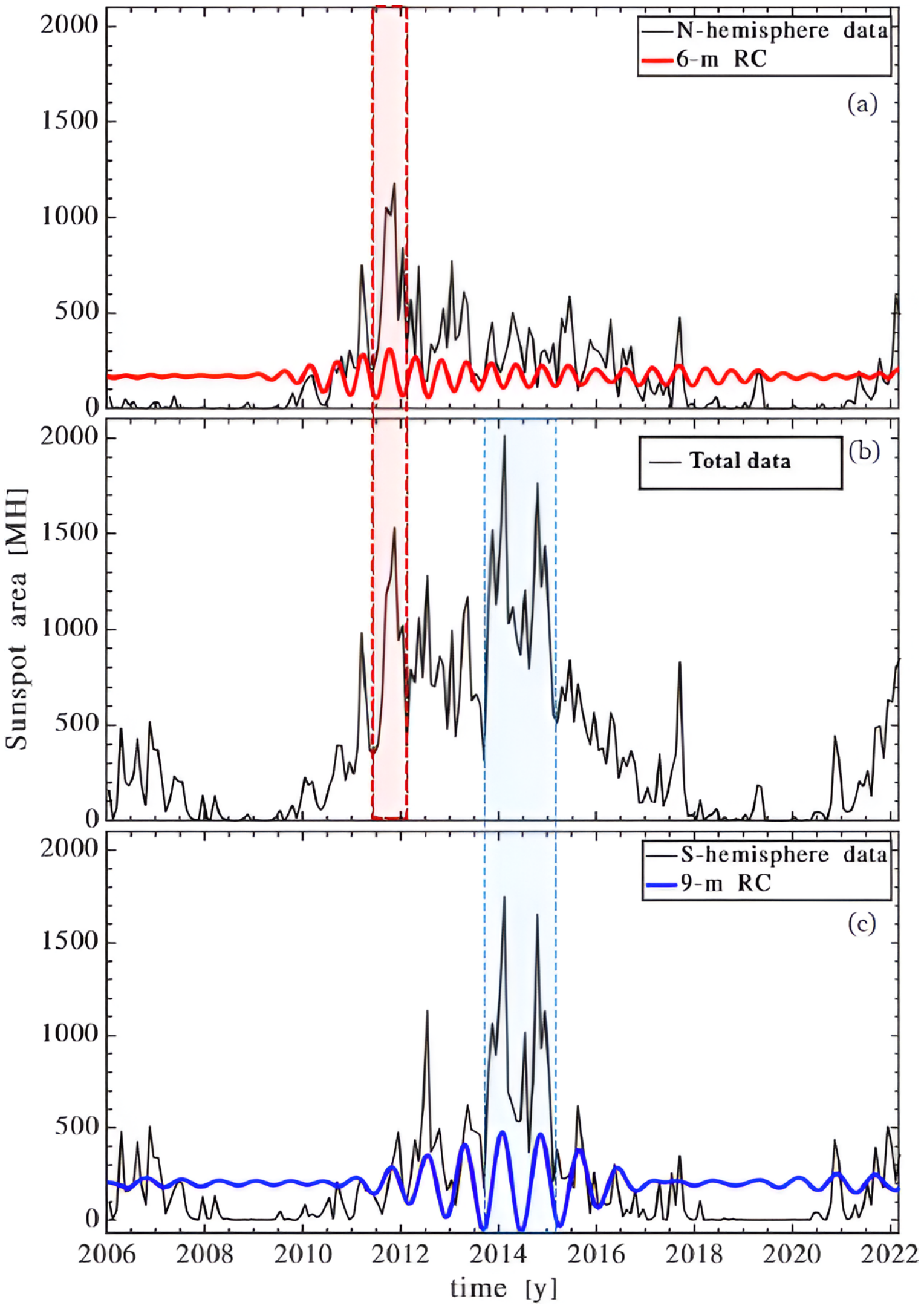

**Figure 5.** Comparison between the full Sun (b), Northern N- (a) and Southern S- (c) Hemispheres sunspot area time series (black curves) sampled every 30 d, to better distinguish the main peaks by reducing the noise in the signal. The NH and SH data are superimposed by the reconstructed ~6 m (red curve in panel a) and ~9 m (blue curve in panel c) components, respectively, obtained by applying the MC-SSA to the 6 d sampled series. The shaded red and blue bars highlight the main peaks in solar activity linked to higher variability in the 6 and 9 m components.

### 3.4. *Monthly variability*

A monthly component, with a period of about 28 d, was also revealed in the scalers series and reconstructed in Figure 3(f). This component can be related to the combination of solar rotation and an inhomogeneous distribution of long-lived solar active regions, such as sunspots, coronal holes, and co-rotating interaction regions (Grieder 2001). The amplitude of this variability component in the scalers reaches its maximum value around solar activity maximum, as results from the comparison of panels (f) and the bottom panel of Figure 3. Furthermore, a higher variability is observed until the beginning of 2018, that is, for the entire duration of the descending phase of the cycle. In fact, while CMEs approximately follow the SN variation (Yashiro et al. 2004), CIRs are more prevalent in the declining phase of the solar cycle (Richardson et al. 2002).

This cycle is linked to the longitudinal asymmetry of the electromagnetic conditions in the heliosphere during one solar rotation. Due to solar differential rotation, the sidereal rotational period of the photosphere is about 25 d at the near-equatorial region and reaches values around 34 d at about ±75° of heliographic latitude. On the other hand, it is important to consider that large-scale coronal magnetic structures rotate more rigidly than the underlying photosphere, that is, much faster at all latitudes and less differentially than the underlying small-scale magnetic structures linked to the photospheric plasma (Mancuso & Giordano 2011; Mancuso et al. 2020), naturally leading to the observed predominance of the observed ~28 d periodicity, corresponding to the so-called synodic rotation period, the apparent rotation period of the Sun as seen from the Earth. This variability has also been found in several neutron monitor data (Modzelewska & Gil 2021; Bazilevskaya 2000; López-Comazzi & Blanco 2022, 2020) and has been detected in the SN time series, as shown in Figure 4(b). Thus, the quasi-periodic modulation of the cosmic-ray flux over 28 days can be mainly attributed to the quasi-rigid rotation of coronal magnetic structures and is also related to the CIRs.

| SSA SIGNIFICANT COMPONENT (99% c.l.) | | | | |
|---|---|---|---|---|
| | SCALERS | TOTAL SUNSPOT AREA | SUNSPOT AREA NORTH | SUNSPOT AREA SOUTH |
| Period | Variance [%] | | | |
| 11 y | 68.2 | 53.7 | 37.8 | 40.7 |
| 1 y | 14.8 | - | - | - |
| ~9 months | 1.0 | 4.0 | - | 6.2 |
| ~6 months | 1.6 | 2.4 | 3.4* | - |
| ~28 d | 2.1 | 7.0 | 8.1 | 12.0 |
| 20 d | 0.4 | 3.0 | 3.5 | - |
| ~14 d | 0.2 | - | - | - |
| *90% c.l. | | | | |
| SIGNAL | ~88% | ~70% | ~53% | ~59% |
| NOISE | ~12% | ~30% | ~47% | ~41% |

**Table 1.** The percentage of the total variance associated with the SSA significant components of the three sunspot area series and the scalers series. The last two rows show the total variance related to signal and noise for each series.

### 3.5. *The 14-day variability*

A significant 14 d oscillation was also revealed in the scaler time series. The reconstruction of this component is shown in Figure 3(h). Higher variability is observed between 2015 and 2019 and around 2008, i.e., in correspondence with declining phases of solar cycles. This is in agreement with the results reported in Ref. (Mursula & Zieger 1996): the largest variability of this oscillation has been found in the late declining phase of the solar cycle in the case of heliospheric variables and around sunspot maxima in the case of solar variables. In the heliospheric case, intervals of large 14 d periodicity are generally attributed to the occurrence of two high-speed solar wind streams approximately 180° apart in solar longitude, per solar rotation. Due to the tilt of the solar quasi-dipolar global magnetic field, such two-stream structures naturally appear if the heliospheric current sheet is narrow and the dipole axis is tilted in relation to the solar rotation axis (Vipindas et al. 2016). Spectral peaks at about half the solar rotation period have also appeared in solar wind studies and geomagnetic activity. It has also been found in neutron monitor data and several solar indices (López-Comazzi & Blanco 2020).

## 4. DISCUSSION AND CONCLUSIONS

Solar activity variations strongly impact the modulation of the flux of low-energy GCRs reaching the Earth. In this work, we have shown that, through the spectral scaler analysis data from the Pierre Auger Observatory, the time variations of the GCR flux related to the activity of the heliosphere at different time scales can be revealed with high accuracy.

Several periodic components have been detected in the 16-year-long scaler rate time series with a 99% c.l. against the null hypothesis of a red noise process. The series has been sampled every 6 days to investigate scales from decadal to a few days. A dominant decadal modulation has been detected, describing most of the signal variance. This reveals an evident anticorrelation with the decadal solar cycle, which, for Solar Cycle 24, lasted 11 years. An observed phase displacement that varies with time has been explained as due to the polarity reversal of the solar magnetic field. An annual oscillation is also present, showing maxima in correspondence with austral winters and minima during austral summers. Its origin is possibly linked to both terrestrial and extraterrestrial causes, such as the temperature effect affecting the muon flux at the surface, the variation of the Earth-Sun distance during the year, and the effect of the asymmetry of the heliospheric magnetic field.

Other shorter-term oscillations with periods of about 9 and 6 months, 28, 20 and 14 days have also been detected. The first two components have been detected in several neutron monitor data and many solar indices. The analysis of the SA series corresponding to the northern and southern hemispheres has revealed the different origins of the two oscillations. The 28-day component shows higher variability in correspondence with both the maximum and the descending phases of Solar Cycle 24. This modulation is caused by the combination of an inhomogeneous distribution of long-lived solar active regions and solar rotation. This periodicity has also been found in the SN time series. The 14-day periodic component is associated with both solar active longitudes and tilted dipole structures. It has also been found in neutron monitor data and several solar indices. Also, this component is observed to be more prominent during the declining phase of solar cycles, presenting a higher variability between 2015 and 2019 and around 2008.

In conclusion, we have shown that Auger scaler data are strongly related to solar activity. The intrinsic very low noise level and the high statistical significance related to the very high count rates also allow a detailed investigation of the GCR flux variations in the heliosphere at different timescales by revealing signals of very low amplitude.

The scaler information is also available in the AugerPrime SD electronics and therefore these studies can be extended by using the dataset beyond 2022.

**Acknowledgements**

The successful installation, commissioning, and operation of the Pierre Auger Observatory would not have been possible without the strong commitment and effort from the technical and administrative staff in Malargüe. We are very grateful to the following agencies and organizations for financial support: Argentina – Comisión Nacional de Energía Atómica; Agencia Nacional de Promoción Científica y Tecnológica (ANPCyT); Consejo Nacional de Investigaciones Científicas y Técnicas (CONICET); Gobierno de la Provincia de Mendoza; Municipalidad de Malargüe; NDM Holdings and Valle Las Leñas; in gratitude for their continuing cooperation over land access; Australia – the Australian Research Council; Belgium – Fonds de la Recherche Scientifique (FNRS); Research Foundation Flanders (FWO), Marie Curie Action of the European Union Grant No. 101107047; Brazil – Conselho Nacional de Desenvolvimento Científico e Tecnológico (CNPq); Financiadora de Estudos e Projetos (FINEP); Fundação de Amparo à Pesquisa do Estado de Rio de Janeiro (FAPERJ); São Paulo Research Foundation (FAPESP) Grants No. 2019/10151-2, No. 2010/07359-6 and No. 1999/05404-3; Ministério da Ciência, Tecnologia, Inovações e Comunicações (MCTIC); Czech Republic – GACR 24-13049S, CAS LQ100102401, MEYS LM2023032, CZ.02.1.01/0.0/0.0/16_013/0001402, CZ.02.1.01/0.0/0.0/18_046/0016010 and CZ.02.1.01/0.0/0.0/17_049/0008422 and CZ.02.01.01/00/22_008/0004632; France – Centre de Calcul IN2P3/CNRS; Centre National de la Recherche Scientifique (CNRS); Conseil Régional Ile-de-France; Département Physique Nucléaire et Corpusculaire (PNC-IN2P3/CNRS); Département Sciences de l'Univers (SDU-INSU/CNRS); Institut Lagrange de Paris (ILP) Grant No. LABEX ANR-10-LABX-63 within the Investissements d'Avenir Programme Grant No. ANR-11-IDEX-0004-02; Germany – Bundesministerium für Bildung und Forschung (BMBF); Deutsche Forschungsgemeinschaft (DFG); Finanzministerium Baden-Württemberg; Helmholtz Alliance for Astroparticle Physics (HAP); Helmholtz-Gemeinschaft Deutscher Forschungszentren (HGF); Ministerium für Kultur und Wissenschaft des Landes Nordrhein-Westfalen; Ministerium für Wissenschaft, Forschung und Kunst des Landes Baden-Württemberg; Italy – Istituto Nazionale di Fisica Nucleare (INFN); Istituto Nazionale di Astrofisica (INAF); Ministero dell'Università e della Ricerca (MUR); CETEMPS Center of Excellence; Ministero degli Affari Esteri (MAE), ICSC Centro Nazionale di Ricerca in High Performance Computing, Big Data and Quantum Computing, funded by European Union NextGenerationEU, reference code CN_00000013; México – Consejo Nacional de Ciencia y Tecnología (CONACYT) No. 167733; Universidad Nacional Autónoma de México (UNAM); PAPIIT DGAPA-UNAM; The Netherlands – Ministry of Education, Culture and Science; Netherlands Organisation for Scientific Research (NWO); Dutch national e-infrastructure with the support of SURF Cooperative; Poland – Ministry of Education and Science, grants No. DIR/WK/2018/11 and 2022/WK/12; National Science Centre, grants No. 2016/22/M/ST9/00198, 2016/23/B/ST9/01635, 2020/39/B/ST9/01398, and 2022/45/B/ST9/02163; Portugal – Portuguese national funds and FEDER funds within Programa Operacional Factores de Competitividade through Fundação para a Ciência e a Tecnologia (COMPETE); Romania – Ministry of Research, Innovation and Digitization, CNCS-UEFISCDI, contract no. 30N/2023 under Romanian National Core Program LAPLAS VII, grant no. PN 23 21 01 02 and project number PN-III-P1-1.1-TE-2021-0924/TE57/2022, within PNCDI III; Slovenia – Slovenian Research Agency, grants P1-0031, P1-0385, I0-0033, N1-0111; Spain – Ministerio de Ciencia e Innovación/Agencia Estatal de Investigación (PID2019-105544GB-I00, PID2022-140510NB-I00 and RYC2019-027017-I), Xunta de Galicia (CIGUS Network of Research Centers, Consolidación 2021 GRC GI-2033, ED431C-2021/22 and ED431F-2022/15), Junta de Andalucía (SOMM17/6104/UGR and P18-FR-4314), and the European Union (Marie Sklodowska-Curie 101065027 and ERDF); USA – Department of Energy, Contracts No. DE-AC02-07CH11359, No. DE-FR02-04ER41300, No. DE-FG02-99ER41107 and No. DE-SC0011689; National Science Foundation, Grant No. 0450696; The Grainger Foundation; Marie Curie-IRSES/EPLANET; European Particle Physics Latin American Network; and UNESCO.

The SSA analyses were performed using the SSA-MTM Toolkit at https://research.aos.ucla.edu/dkondras/ssa/form.html. Daily mean sunspot numbers come from the source: WDC-SILSO, Royal Observatory of Belgium, Brussels, and these can be downloaded from https://www.sidc.be/silso/.

# APPENDIX

## A. MONTE CARLO SINGULAR SPECTRUM ANALYSIS

The Singular Spectrum Analysis (SSA) is a non-parametric spectral method that allows to efficiently extract the deterministic components of a time series from noise (Vautard & Ghil 1989; Vautard et al. 1992; Ghil et al. 2002). It uses data-adaptive filters to separate the time series into statistically independent components that can be classified as oscillatory patterns modulated in amplitude and phase.

The SSA methodology applied on a time series $x(n)$ consists of three basic steps:

- embedding the time series of length $N$ in a vector space of proper dimension $M$;
- computing the $M \times M$ lag-covariance matrix $C_D$ of the data;
- diagonalizing the matrix $C_D$ to extract its eigenvectors $E_k$ and corresponding eigenvalues $\lambda_k$, with $k = 1, \dots M$.

The lag-covariance matrix $C_D$ can be estimated as

$$C_D = \frac{1}{N'} D^{\mathrm{T}} D, \tag{A1}$$

where $N' = N - M + 1$ is the embedding dimension, and $D$ is the trajectory matrix, defined as

$$D = \begin{pmatrix} x(1) & x(2) & \dots & x(M) \\ x(2) & x(3) & \dots & x(M+1) \\ \vdots & \vdots & \ddots & \vdots \\ x(N') & x(N'+1) & \dots & x(N) \end{pmatrix}. \tag{A2}$$

The diagonalization of the matrix $C_D$ yields the diagonal matrix $\Lambda_D = E_D^{\mathrm{T}} C_D E_D$, where $\Lambda_D = \mathrm{diag}(\lambda_1, \lambda_2, \lambda_3, \dots, \lambda_M)$, with $\lambda_1 > \lambda_2 > \lambda_3 > \dots > \lambda_M > 0$, and $E_D$ is the $M \times M$ matrix having the corresponding eigenvectors $E_k$, $k = 1, \dots M$, as its columns.

For each $E_k$, also known as Empirical Orthogonal Functions (EOFs), the $k$-th principal component (PC) is constructed, representing a time series of length $N'$ obtained by projecting the original time series on the eigenvector $E_k$,

$$P_k(t') = \sum_{j=1}^{M} x(t' + j) \, E_k(j). \tag{A3}$$

The corresponding eigenvalue $\lambda_k$ describes its variance, which can be interpreted as the percentage of the time series $x(n)$ described by the $k$-th component.

Given a subset of eigenvalues, it is possible to extract a time series of length $N$ by combining the corresponding PCs. These time series, called Reconstructed Components (RCs), capture the variability associated with the eigenvalues of interest and are estimated as

$$R_k(t) = \frac{1}{M_t} \sum_{k \in K} \sum_{j=L_t}^{U_t} P_k(t - j) \, E_k(j). \tag{A4}$$

The values of the normalization factor $M_t$ and the lower and upper bound of summation $L_t$ and $U_t$ differ between the central part of the time series and its endpoints (Vautard et al. 1992; Ghil et al. 2002),

$$(M_t, L_t, U_t) = \begin{cases} (\frac{1}{t}, 1, t) & ;\ 1 \leqslant t \leqslant M - 1, \\ (\frac{1}{M}, 1, M) & ;\ M \leqslant t \leqslant N', \\ (\frac{1}{N-t+1}, t - N + M, M) & ;\ N' + 1 \leqslant t \leqslant N. \end{cases} \tag{A5}$$

**The Pierre Auger Collaboration**

PIERRE AUGER OBSERVATORY

A. Abdul Halim[13], P. Abreu[70], M. Aglietta[53,51], I. Allekotte[1], K. Almeida Cheminant[78,77], A. Almela[7,12], R. Aloisio[44,45], J. Alvarez-Muñiz[76], A. Ambrosone[44], J. Ammerman Yebra[76], G.A. Anastasi[57,46], L. Anchordoqui[83], B. Andrada[7], L. Andrade Dourado[44,45], S. Andringa[70], L. Apollonio[58,48], C. Aramo[49], E. Arnone[62,51], J.C. Arteaga Velázquez[66], P. Assis[70], G. Avila[11], E. Avocone[56,45], A. Bakalova[31], F. Barbato[44,45], A. Bartz Mocellin[82], J.A. Bellido[13], C. Berat[35], M.E. Bertaina[62,51], M. Bianciotto[62,51], P.L. Biermann[a], V. Binet[5], K. Bismark[38,7], T. Bister[77,78], J. Biteau[36,i], J. Blazek[31], J. Blümer[40], M. Boháčová[31], D. Boncioli[56,45], C. Bonifazi[8], L. Bonneau Arbeletche[22], N. Borodai[68], J. Brack[f], P.G. Brichetto Orchera[7], F.L. Briechle[41], A. Bueno[75], S. Buitink[15], M. Buscemi[46,57], M. Büsken[38,7], A. Bwembya[77,78], K.S. Caballero-Mora[65], S. Cabana-Freire[76], L. Caccianiga[58,48], F. Campuzano[6], J. Caraça-Valente[82], R. Caruso[57,46], A. Castellina[53,51], F. Catalani[19], G. Cataldi[47], L. Cazon[76], M. Cerda[10], B. Čermáková[40], A. Cermenati[44,45], J.A. Chinellato[22], J. Chudoba[31], L. Chytka[32], R.W. Clay[13], A.C. Cobos Cerutti[6], R. Colalillo[59,49], R. Conceição[70], A. Condorelli[36], G. Consolati[48,54], M. Conte[55,47], F. Convenga[56,45], D. Correia dos Santos[27], P.J. Costa[70], C.E. Covault[81], M. Cristinziani[43], C.S. Cruz Sanchez[3], S. Dasso[4,2], K. Daumiller[40], B.R. Dawson[13], R.M. de Almeida[27], E.-T. de Boone[43], B. de Errico[27], J. de Jesús[7,40], S.J. de Jong[77,78], J.R.T. de Mello Neto[27], I. De Mitri[44,45], J. de Oliveira[18], D. de Oliveira Franco[42], F. de Palma[55,47], V. de Souza[20], E. De Vito[55,47], A. Del Popolo[57,46], O. Deligny[33], N. Denner[31], L. Deval[40,7], A. di Matteo[51], C. Dobrigkeit[22], J.C. D'Olivo[67], L.M. Domingues Mendes[16,70], Q. Dorosti[43], J.C. dos Anjos[16], R.C. dos Anjos[26], J. Ebr[31], F. Ellwanger[40], M. Emam[77,78], R. Engel[38,40], I. Epicoco[55,47], M. Erdmann[41], A. Etchegoyen[7,12], C. Evoli[44,45], H. Falcke[77,79,78], G. Farrar[85], A.C. Fauth[22], T. Fehler[43], F. Feldbusch[39], A. Fernandes[70], B. Fick[84], J.M. Figueira[7], P. Filip[38,7], A. Filipčič[74,73], T. Fitoussi[40], B. Flaggs[87], T. Fodran[77], M. Freitas[70], T. Fujii[86,h], A. Fuster[7,12], C. Galea[77], B. García[6], C. Gaudu[37], P.L. Ghia[33], U. Giaccari[47], F. Gobbi[10], F. Gollan[7], G. Golup[1], M. Gómez Berisso[1], P.F. Gómez Vitale[11], J.P. Gongora[11], J.M. González[1], N. González[7], D. Góra[68], A. Gorgi[53,51], M. Gottowik[40], F. Guarino[59,49], G.P. Guedes[23], E. Guido[43], L. Gülzow[40], S. Hahn[38], P. Hamal[31], M.R. Hampel[7], P. Hansen[3], V.M. Harvey[13], A. Haungs[40], T. Hebbeker[41], C. Hojvat[d], J.R. Hörandel[77,78], P. Horvath[32], M. Hrabovský[32], T. Huege[40,15], A. Insolia[57,46], P.G. Isar[72], P. Janecek[31], V. Jilek[31], K.-H. Kampert[37], B. Keilhauer[40], A. Khakurdikar[77], V.V. Kizakke Covilakam[7,40], H.O. Klages[40], M. Kleifges[39], J. Köhler[40], F. Krieger[41], M. Kubatova[31], N. Kunka[39], B.L. Lago[17], N. Langner[41], M.A. Leigui de Oliveira[25], Y. Lema-Capeans[76], A. Letessier-Selvon[34], I. Lhenry-Yvon[33], L. Lopes[70], J.P. Lundquist[73], A. Machado Payeras[22], M. Mallamaci[60,46], D. Mandat[31], B.C. Manning[13], P. Mantsch[d], F.M. Mariani[58,48], A.G. Mariazzi[3], I.C. Mariş[14], G. Marsella[60,46], D. Martello[55,47], S. Martinelli[40,7], M.A. Martins[76], H.-J. Mathes[40], J. Matthews[g], G. Matthiae[61,50], E. Mayotte[82], S. Mayotte[82], P.O. Mazur[d], G. Medina-Tanco[67], J. Meinert[37], D. Melo[7], A. Menshikov[39], C. Merx[40], S. Michal[31], M.I. Micheletti[5], L. Miramonti[58,48], M. Mogarkar[68], S. Mollerach[1], F. Montanet[35], L. Morejon[37], K. Mulrey[77,78], R. Mussa[51], W.M. Namasaka[37], S. Negi[31], L. Nellen[67], K. Nguyen[84], G. Nicora[9], M. Niechciol[43], D. Nitz[84], D. Nosek[30], A. Novikov[87], V. Novotny[30], L. Nožka[32], A. Nucita[55,47], L.A. Núñez[29], J. Ochoa[7], C. Oliveira[20], L. Östman[31], M. Palatka[31], J. Pallotta[9], S. Panja[31], G. Parente[76], T. Paulsen[37], J. Pawlowsky[37], M. Pech[31], J. Pękala[68], R. Pelayo[64], V. Pelgrims[14], L.A.S. Pereira[24], E.E. Pereira Martins[38,7], C. Pérez Bertolli[7,40], L. Perrone[55,47], S. Petrera[44,45], C. Petrucci[56], T. Pierog[40], M. Pimenta[70], M. Platino[7], B. Pont[77], M. Pourmohammad Shahvar[60,46], P. Privitera[86], M. Prouza[31], K. Pytel[69], S. Querchfeld[37], J. Rautenberg[37], D. Ravignani[7], J.V. Reginatto Akim[22], A. Reuzki[41], J. Ridky[31], F. Riehn[76,j], M. Risse[43], V. Rizi[56,45], E. Rodriguez[7,40], G. Rodriguez Fernandez[50], J. Rodriguez Rojo[11], M.J. Roncoroni[7], S. Rossoni[42], M. Roth[40], E. Roulet[1], A.C. Rovero[4], A. Saftoiu[71], M. Saharan[77], F. Salamida[56,45], H. Salazar[63], G. Salina[50], P. Sampathkumar[40], N. San Martin[82], J.D. Sanabria Gomez[29], F. Sánchez[7], E.M. Santos[21], E. Santos[31], F. Sarazin[82], R. Sarmento[70], R. Sato[11], P. Savina[44,45], V. Scherini[55,47], H. Schieler[40], M. Schimassek[33], M. Schimp[37], D. Schmidt[40], O. Scholten[15,b], H. Schoorlemmer[77,78], P. Schovánek[31], F.G. Schröder[87,40], J. Schulte[41], T. Schulz[40,7], S.J. Sciutto[3], M. Scornavacche[7,40], A. Sedoski[7], A. Segreto[52,46], S. Sehgal[37], S.U. Shivashankara[73], G. Sigl[42], K. Simkova[15,14], F. Simon[39], R. Šmída[86], P. Sommers[e], R. Squartini[10], M. Stadelmaier[40,48,58], S. Stanič[73], J. Stasielak[68], P. Stassi[35], S. Strähnz[38], M. Straub[41], T. Suomijärvi[36], A.D. Supanitsky[7], Z. Svozilikova[31], Z. Szadkowski[69], F. Tairli[13], A. Tapia[28], C. Taricco[62,51], C. Timmermans[78,77], O. Tkachenko[31], P. Tobiska[31], C.J. Todero Peixoto[19], B. Tomé[70], A. Travaini[10], P. Travnicek[31], M. Tueros[3], M. Unger[40], R. Uzeiroska[37], L. Vaclavek[32], M. Vacula[32], I. Vaiman[44,45], J.F. Valdés Galicia[67], L. Valore[59,49], E. Varela[63], V. Vašíčková[37], A. Vásquez-Ramírez[29], D. Veberič[40], I.D. Vergara Quispe[3], S. Verpoest[87], V. Verzi[50], J. Vicha[31], J. Vink[80], S. Vorobiov[73], J.B. Vuta[31], C. Watanabe[27], A.A. Watson[c], A. Weindl[40], M. Weitz[37], L. Wiencke[82], H. Wilczyński[68], D. Wittkowski[37], B. Wundheiler[7], B. Yue[37], A. Yushkov[31], E. Zas[76], D. Zavrtanik[73,74], M. Zavrtanik[74,73]

[1] Centro Atómico Bariloche and Instituto Balseiro (CNEA-UNCuyo-CONICET), San Carlos de Bariloche, Argentina
[2] Departamento de Física and Departamento de Ciencias de la Atmósfera y los Océanos, FCEyN, Universidad de Buenos Aires and CONICET, Buenos Aires, Argentina
[3] IFLP, Universidad Nacional de La Plata and CONICET, La Plata, Argentina
[4] Instituto de Astronomía y Física del Espacio (IAFE, CONICET-UBA), Buenos Aires, Argentina
[5] Instituto de Física de Rosario (IFIR) – CONICET/U.N.R. and Facultad de Ciencias Bioquímicas y Farmacéuticas U.N.R., Rosario, Argentina
[6] Instituto de Tecnologías en Detección y Astropartículas (CNEA, CONICET, UNSAM), and Universidad Tecnológica Nacional – Facultad Regional Mendoza (CONICET/CNEA), Mendoza, Argentina
[7] Instituto de Tecnologías en Detección y Astropartículas (CNEA, CONICET, UNSAM), Buenos Aires, Argentina
[8] International Center of Advanced Studies and Instituto de Ciencias Físicas, ECyT-UNSAM and CONICET, Campus Miguelete – San Martín, Buenos Aires, Argentina
[9] Laboratorio Atmósfera – Departamento de Investigaciones en Láseres y sus Aplicaciones – UNIDEF (CITEDEF-CONICET), Argentina
[10] Observatorio Pierre Auger, Malargüe, Argentina
[11] Observatorio Pierre Auger and Comisión Nacional de Energía Atómica, Malargüe, Argentina
[12] Universidad Tecnológica Nacional – Facultad Regional Buenos Aires, Buenos Aires, Argentina
[13] University of Adelaide, Adelaide, S.A., Australia
[14] Université Libre de Bruxelles (ULB), Brussels, Belgium
[15] Vrije Universiteit Brussels, Brussels, Belgium
[16] Centro Brasileiro de Pesquisas Fisicas, Rio de Janeiro, RJ, Brazil
[17] Centro Federal de Educação Tecnológica Celso Suckow da Fonseca, Petropolis, Brazil
[18] Instituto Federal de Educação, Ciência e Tecnologia do Rio de Janeiro (IFRJ), Brazil
[19] Universidade de São Paulo, Escola de Engenharia de Lorena, Lorena, SP, Brazil
[20] Universidade de São Paulo, Instituto de Física de São Carlos, São Carlos, SP, Brazil
[21] Universidade de São Paulo, Instituto de Física, São Paulo, SP, Brazil
[22] Universidade Estadual de Campinas (UNICAMP), IFGW, Campinas, SP, Brazil
[23] Universidade Estadual de Feira de Santana, Feira de Santana, Brazil
[24] Universidade Federal de Campina Grande, Centro de Ciencias e Tecnologia, Campina Grande, Brazil
[25] Universidade Federal do ABC, Santo André, SP, Brazil
[26] Universidade Federal do Paraná, Setor Palotina, Palotina, Brazil
[27] Universidade Federal do Rio de Janeiro, Instituto de Física, Rio de Janeiro, RJ, Brazil
[28] Universidad de Medellín, Medellín, Colombia
[29] Universidad Industrial de Santander, Bucaramanga, Colombia
[30] Charles University, Faculty of Mathematics and Physics, Institute of Particle and Nuclear Physics, Prague, Czech Republic
[31] Institute of Physics of the Czech Academy of Sciences, Prague, Czech Republic
[32] Palacky University, Olomouc, Czech Republic
[33] CNRS/IN2P3, IJCLab, Université Paris-Saclay, Orsay, France
[34] Laboratoire de Physique Nucléaire et de Hautes Energies (LPNHE), Sorbonne Université, Université de Paris, CNRS-IN2P3, Paris, France
[35] Univ. Grenoble Alpes, CNRS, Grenoble Institute of Engineering Univ. Grenoble Alpes, LPSC-IN2P3, 38000 Grenoble, France
[36] Université Paris-Saclay, CNRS/IN2P3, IJCLab, Orsay, France
[37] Bergische Universität Wuppertal, Department of Physics, Wuppertal, Germany
[38] Karlsruhe Institute of Technology (KIT), Institute for Experimental Particle Physics, Karlsruhe, Germany
[39] Karlsruhe Institute of Technology (KIT), Institut für Prozessdatenverarbeitung und Elektronik, Karlsruhe, Germany
[40] Karlsruhe Institute of Technology (KIT), Institute for Astroparticle Physics, Karlsruhe, Germany
[41] RWTH Aachen University, III. Physikalisches Institut A, Aachen, Germany
[42] Universität Hamburg, II. Institut für Theoretische Physik, Hamburg, Germany
[43] Universität Siegen, Department Physik – Experimentelle Teilchenphysik, Siegen, Germany
[44] Gran Sasso Science Institute, L'Aquila, Italy
[45] INFN Laboratori Nazionali del Gran Sasso, Assergi (L'Aquila), Italy
[46] INFN, Sezione di Catania, Catania, Italy
[47] INFN, Sezione di Lecce, Lecce, Italy
[48] INFN, Sezione di Milano, Milano, Italy
[49] INFN, Sezione di Napoli, Napoli, Italy
[50] INFN, Sezione di Roma "Tor Vergata", Roma, Italy
[51] INFN, Sezione di Torino, Torino, Italy
[52] Istituto di Astrofisica Spaziale e Fisica Cosmica di Palermo (INAF), Palermo, Italy
[53] Osservatorio Astrofisico di Torino (INAF), Torino, Italy

[54] Politecnico di Milano, Dipartimento di Scienze e Tecnologie Aerospaziali , Milano, Italy
[55] Università del Salento, Dipartimento di Matematica e Fisica “E. De Giorgi”, Lecce, Italy
[56] Università dell’Aquila, Dipartimento di Scienze Fisiche e Chimiche, L’Aquila, Italy
[57] Università di Catania, Dipartimento di Fisica e Astronomia “Ettore Majorana“, Catania, Italy
[58] Università di Milano, Dipartimento di Fisica, Milano, Italy
[59] Università di Napoli “Federico II”, Dipartimento di Fisica “Ettore Pancini”, Napoli, Italy
[60] Università di Palermo, Dipartimento di Fisica e Chimica ”E. Segrè”, Palermo, Italy
[61] Università di Roma “Tor Vergata”, Dipartimento di Fisica, Roma, Italy
[62] Università Torino, Dipartimento di Fisica, Torino, Italy
[63] Benemérita Universidad Autónoma de Puebla, Puebla, México
[64] Unidad Profesional Interdisciplinaria en Ingeniería y Tecnologías Avanzadas del Instituto Politécnico Nacional (UPIITA-IPN), México, D.F., México
[65] Universidad Autónoma de Chiapas, Tuxtla Gutiérrez, Chiapas, México
[66] Universidad Michoacana de San Nicolás de Hidalgo, Morelia, Michoacán, México
[67] Universidad Nacional Autónoma de México, México, D.F., México
[68] Institute of Nuclear Physics PAN, Krakow, Poland
[69] University of Łódź, Faculty of High-Energy Astrophysics,Łódź, Poland
[70] Laboratório de Instrumentação e Física Experimental de Partículas – LIP and Instituto Superior Técnico – IST, Universidade de Lisboa – UL, Lisboa, Portugal
[71] “Horia Hulubei” National Institute for Physics and Nuclear Engineering, Bucharest-Magurele, Romania
[72] Institute of Space Science, Bucharest-Magurele, Romania
[73] Center for Astrophysics and Cosmology (CAC), University of Nova Gorica, Nova Gorica, Slovenia
[74] Experimental Particle Physics Department, J. Stefan Institute, Ljubljana, Slovenia
[75] Universidad de Granada and C.A.F.P.E., Granada, Spain
[76] Instituto Galego de Física de Altas Enerxías (IGFAE), Universidade de Santiago de Compostela, Santiago de Compostela, Spain
[77] IMAPP, Radboud University Nijmegen, Nijmegen, The Netherlands
[78] Nationaal Instituut voor Kernfysica en Hoge Energie Fysica (NIKHEF), Science Park, Amsterdam, The Netherlands
[79] Stichting Astronomisch Onderzoek in Nederland (ASTRON), Dwingeloo, The Netherlands
[80] Universiteit van Amsterdam, Faculty of Science, Amsterdam, The Netherlands
[81] Case Western Reserve University, Cleveland, OH, USA
[82] Colorado School of Mines, Golden, CO, USA
[83] Department of Physics and Astronomy, Lehman College, City University of New York, Bronx, NY, USA
[84] Michigan Technological University, Houghton, MI, USA
[85] New York University, New York, NY, USA
[86] University of Chicago, Enrico Fermi Institute, Chicago, IL, USA
[87] University of Delaware, Department of Physics and Astronomy, Bartol Research Institute, Newark, DE, USA

—

[a] Max-Planck-Institut für Radioastronomie, Bonn, Germany
[b] also at Kapteyn Institute, University of Groningen, Groningen, The Netherlands
[c] School of Physics and Astronomy, University of Leeds, Leeds, United Kingdom
[d] Fermi National Accelerator Laboratory, Fermilab, Batavia, IL, USA
[e] Pennsylvania State University, University Park, PA, USA
[f] Colorado State University, Fort Collins, CO, USA
[g] Louisiana State University, Baton Rouge, LA, USA
[h] now at Graduate School of Science, Osaka Metropolitan University, Osaka, Japan
[i] Institut universitaire de France (IUF), France
[j] now at Technische Universität Dortmund and Ruhr-Universität Bochum, Dortmund and Bochum, Germany